\documentclass{article}

\usepackage[final]{neurips_2022}

\usepackage{amsmath,amssymb}
\usepackage[utf8]{inputenc} 
\usepackage[T1]{fontenc}    
\usepackage{hyperref}       
\usepackage{url}            
\usepackage{booktabs}       
\usepackage{amsfonts}       
\usepackage{nicefrac}       
\usepackage{microtype}      
\usepackage{xcolor}         
\usepackage{multirow}
\usepackage{graphicx}
\usepackage{ifluatex}
\usepackage{dirtytalk}
\usepackage{bm}
\usepackage{tikz}
\usepackage[numbers, sort, comma, square]{natbib}

\usepackage[margins]{trackchanges}

\usepackage{floatrow}
\newfloatcommand{capbtabbox}{table}[][\FBwidth]

\newcommand{\blu}[1]{{\color{blue}{#1}}}

\title{Utilizing Mutations to Evaluate\\Interpretability of Neural Networks on Genomic Data}

\author{%
Utku Ozbulak\thanks{Correspondence: \texttt{utku.ozbulak@ugent.be}}\\
Ghent University\\
\And
Solha Kang\thanks{Work done during an internship at Ghent University.}\\
University of Edinburgh\\
\And
Jasper Zuallaert\\
Ghent University\\
\AND
Stephen Depuydt\\
Ghent University\\
\And
Joris Vankerschaver\\
Ghent University\\
}
\begin{document}

\maketitle

\begin{abstract}
Even though deep neural networks (DNNs) achieve state-of-the-art results for a number of problems involving genomic data, getting DNNs to explain their decision-making process has been a major challenge due to their black-box nature. One way to get DNNs to explain their reasoning for prediction is via attribution methods which are assumed to highlight the parts of the input that contribute to the prediction the most. Given the existence of numerous attribution methods and a lack of quantitative results on the fidelity of those methods, selection of an attribution method for sequence-based tasks has been mostly done qualitatively. In this work, we take a step towards identifying the most faithful attribution method by proposing a computational approach that utilizes point mutations. Providing quantitative results on seven popular attribution methods, we find Layerwise Relevance Propagation (LRP) to be the most appropriate one for translation initiation, with LRP identifying two important biological features for translation: the integrity of Kozak sequence as well as the detrimental effects of premature stop codons.
\end{abstract}

\vspace{-1.5em}

\section{Introduction}
\label{sec:introduction}
\vspace{-0.25em}

\begin{figure}
\begin{tikzpicture}
\node[inner sep=0pt] (org) at (0,0)
    {\includegraphics[width=.15\textwidth]{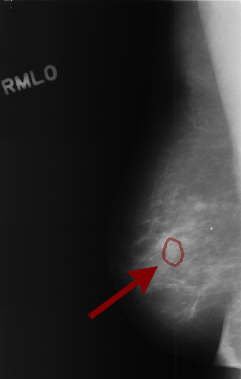}};
\node[inner sep=0pt] (vis1) at (3,1.25)
    {\includegraphics[width=.1\textwidth]{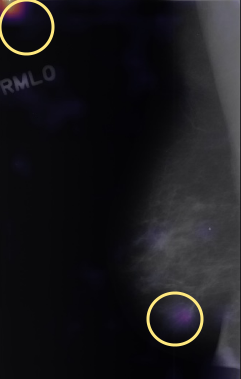}};
\node[inner sep=0pt] (vis2) at (4.6,0)
    {\includegraphics[width=.1\textwidth]{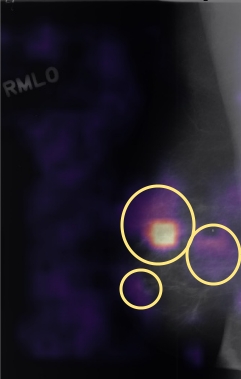}};
\node[inner sep=0pt] (vis3) at (3,-1.25)
    {\includegraphics[width=.1\textwidth]{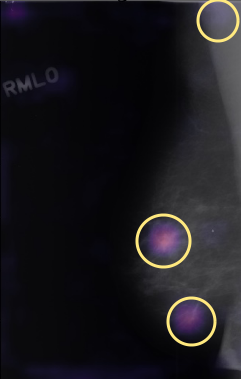}};
\draw[->,thick] (org.east) -- (vis1.west);
\draw[->,thick] (org.east) -- (vis2.west);
\draw[->,thick] (org.east) -- (vis3.west);
\node at (2,-2.75) (CV) {(a) Images};

\node at (9,1) (seq) {\scriptsize \texttt{ACTGGGCTATGGCATATC...GTATGCCCATGACGATACGAT}};
\node[inner sep=0pt] (seq_vis) at (9,0)
    {\includegraphics[width=.4\textwidth]{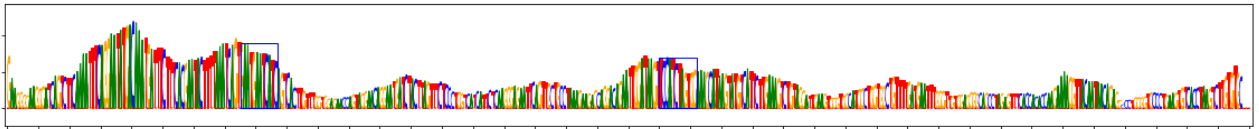}};Genomics
\node[inner sep=0pt] (seq_vis_top) at (9,-0.75)
    {\includegraphics[width=.4\textwidth]{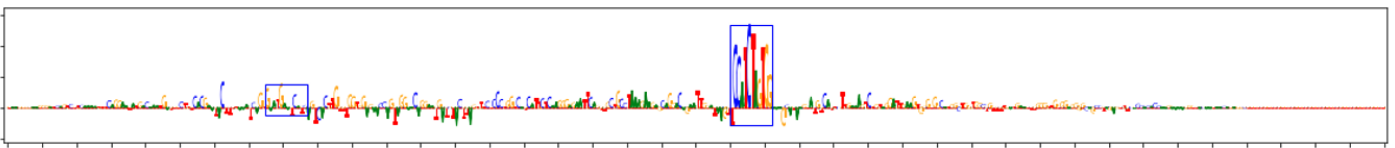}};
\node[inner sep=0pt] (seq_vis_top) at (9,-1.5)
    {\includegraphics[width=.4\textwidth]{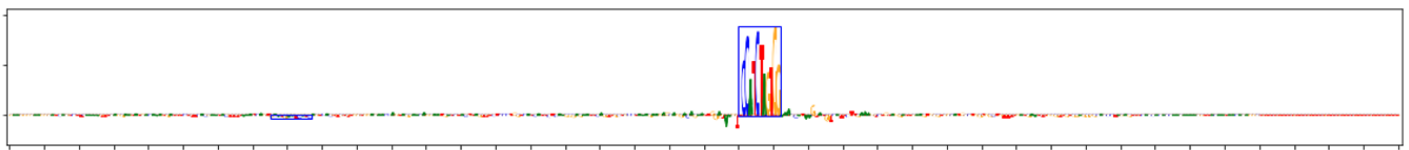}};
\draw[->,thick] (seq.south) -- (seq_vis.north);
\node at (9.25,-2.75) (CV) {(a) Genomic data};
\end{tikzpicture}
\caption{An illustration of conflicting attribution maps for (left) breast cancer detection using images~\cite{wu2021deepminer} and (right) motif detection using genomic data~\cite{prakash2022towards}.}
\label{fig:cv_vs_gen}
\end{figure}
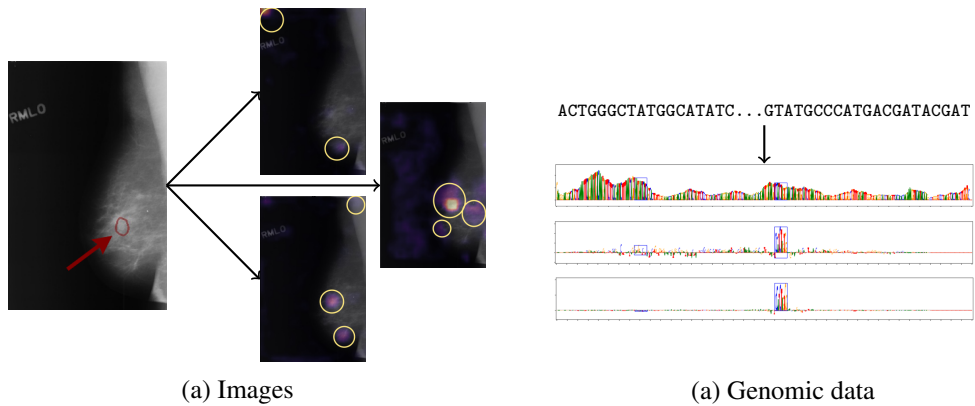

Advancements in machine learning, and deep learning in particular, have contributed to many discoveries in computational biology in recent years, perhaps the most famous research effort being AlphaFold for protein structure prediction~\cite{jumper2021highly}. Nevertheless, using DNNs to uncover highly sought-after biologically relevant features remains a non-trivial task~\cite{ching2018opportunities}. To accomplish this task and to get DNNs to explain their complex reasoning, numerous attribution (also called explainability and interpretability) methods have been employed, with most of these attribution methods originating from the field of computer vision~\cite{LRP_2}. Unfortunately, even on image data these methods often disagree with each other (see Figure~\ref{fig:cv_vs_gen}). Furthermore, recent research efforts show that the usage of attribution methods may lead to misleading results due to attribution methods not generalizing well or making unjustified causal interpretations~\cite{molnar2022general,zhang2020interpretable,rudin2019stop}.




Given the evidence on the instability of attribution methods as well as the availability of a number of unique attribution methods, how can one go about picking the most appropriate attribution method for interpreting DNNs trained with genomic data? A popular method to evaluate the fidelity of attributions for genomic data has been via the usage of synthetic data~\cite{koo2019robust,koo2021improving,prakash2022towards}. Synthetic data in the context of genomics refers to the creation of sequences that contain certain (known) features, such as motifs. Then, the evaluation of attribution methods is performed via observing whether they detect these implanted features or not~\cite{prakash2022towards}. In most cases, synthetic data is preferred over genuine data because it can be modified at will, so that specific, quantitative evaluations can be made. On the other hand, measuring the generalization of observations from synthetic to real data is not straightforward and, in many cases, an open question. Unlike previous research efforts that employ synthetic data, we will demonstrate in this work the usefulness of point mutations in genuine data to evaluate the interpretability of DNN methods. 

Point mutations, in which a single nucleotide is mutated into another one, have a long history in molecular biology in identifying the genetic basis of inherited disases. Although in this context such mutations have been explored mostly in vitro in the past, CRISPR-Cas9 based genome editing enabled an easier and swifter approach~\cite{crispr_mutation2}. Likewise, these mutations are also employed for computational methods to decipher gene function and regulation~\cite{de2002mutation,zhu2020machine}. Taking inspiration from these efforts, we employ point mutations to understand the predictions of DNNs.

Employing the relatively well-documented genomics problem of translation initiation site (TIS) detection, we attempt to make a quantitative evaluation of seven popular attribution methods (\texttt{DeepLift}~\cite{deeplift_1,deeplift_2}, \texttt{Integrated Gradients}~\cite{integrated_grads}, \texttt{LRP}~\cite{LRP_1}, \texttt{Gradient Shap}~\cite{GradientShap}, \texttt{Guided Backpropagation}~\cite{guided_backprop}, \texttt{Kernel Shap}~\cite{GradientShap}, and \texttt{Deconvolution}~\cite{deconvolution}). With large-scale experiments, we not only reveal large discrepancies between attribution outcomes, we also show how certain attributions can correctly identify well-known biological features.

\vspace{-1em}
\section{Experimental Settings}
\label{sec:experimental_setting}
\vspace{-0.5em}

\textbf{Translation initiation site detection}\,\textendash\,A cornerstone of the central dogma of molecular biology is translation, which denotes the synthesis of appropriate proteins from mRNA molecules~\cite{riboseq,kozak1978eucaryotic}. Protein synthesis is principally regulated at the initiation stage, with initiation referring to the assembly of the 80S ribosome in eukaryotes at the start codon with the help of a number of eukaryote initiation factors~\cite{jackson2010mechanism,haberle2018eukaryotic}. As such, precise identification of the translation initiation site on mRNAs is crucial for accurate protein synthesis~\cite{kozak1978eucaryotic}. TIS detection is a complex problem which found a number of well-recognized solutions that leverage DNNs in recent years~\cite{TITER,Neurotis,zuallaert2018tisrover}. Moreover, the availability of public datasets also makes it an attractive problem for model benchmarking. And finally, the biological processes regulating this process on genome-level are relatively well understood compared to a number of other sequencing problems~\cite{jackson2010mechanism}. Thanks to the properties above, we select TIS-detection to evaluate the faithfulness of attribution methods.

\textbf{Data}\,\textendash\,We make use of the \texttt{Human-TIS}, \texttt{CCDS}, and \texttt{Chromosome-21} TIS detection datasets published by~\cite{saeys2007translation,chen2014itis} where the sequences in these datasets are gathered from human DNA sequences (see the supplementary material for details). TIS-detection is a $2$-class classification problem with inputs composed of various permutations of four base pairs (bp) (\texttt{A},\texttt{C}, \texttt{T}, \texttt{G}). The datasets contain sequences consisting of $203$ bp with $60$ bp in the 5' untranslated region (UTR) and $140$ bp downstream of the canonical translation initiation site (\blu{\texttt{ATG}}). A unique property of these data is the codon starting at position $61$, which denotes the start of translation. We assign an index of $+1$ to \texttt{A} in \texttt{ATG} at position $61$ and we label upstream (UTR) and downstream (coding region) nucleotides with decreasing and increasing numbers, respectively.



\textbf{Model}\,\textendash\,We use the TISRover architecture described in the work of~\citet{zuallaert2018tisrover} which achieves state-of-the-art results on this data. Detailed results on the model performance as well as the training routine are provided in the supplementary material.

\textbf{Attribution methods}\,\textendash\,As mentioned in Section~\ref{sec:introduction}, we evaluate seven commonly used attribution methods given in Section~\ref{sec:introduction}. For the implementation of these methods we use the PyTorch Captum library~\cite{kokhlikyan2020captum}. More details for these attribution methods can be found in their respective paper as well as the comprehensive works of~\cite{LRP_2,LRP_3}.

\begin{figure}[t!]
\centering
\includegraphics[width=\textwidth]{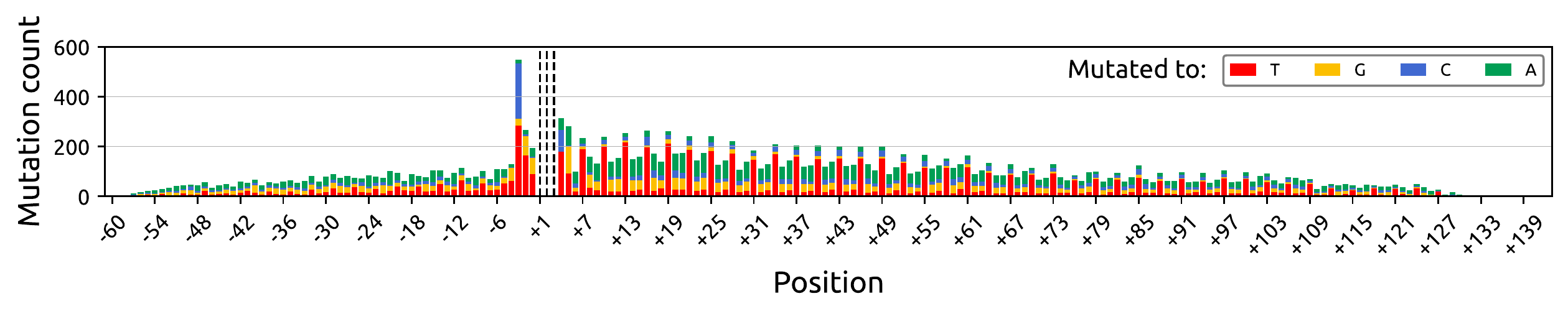}
\vspace{-1.25em}
\includegraphics[width=\textwidth]{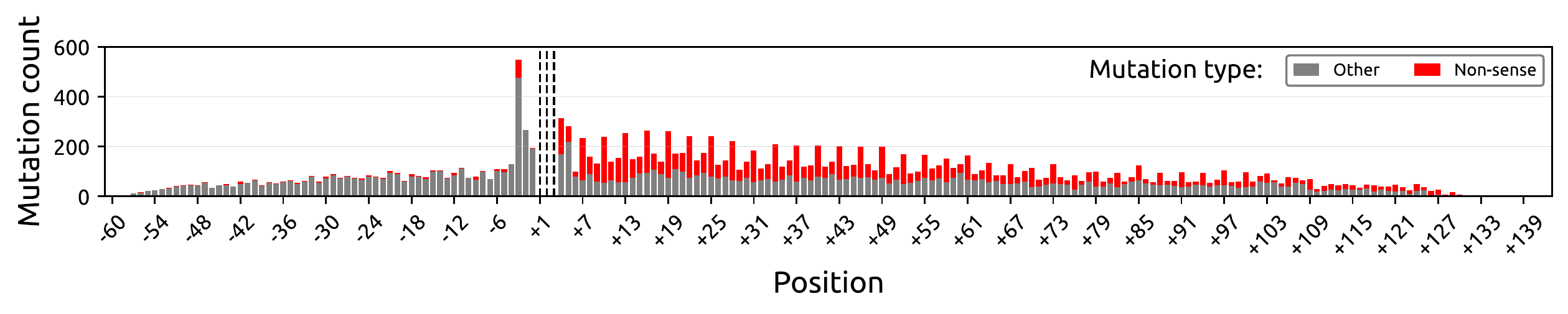}
\caption{(top) Histogram of mutations per position that, when performed, change the prediction made by the model from TIS-positive to TIS-negative and (bottom) distribution of mutations when separated into non-sense and others.}
\label{fig:mut_nuc_origin}
\end{figure}

\vspace{-1em}
\section{Experimental Results}
\label{sec:mutations_and_evaluation}

\vspace{-0.5em}
\subsection{Mutations}
\label{sec:mutations}

Given a TIS-positive sequence consisting of $203$ bp, we create $600$ newly mutated sequences by transforming bp at each position, one at a time, with the other three bases for $200$ positions (not mutating the \texttt{ATG} at position $61$). Our goal in performing this task is to analyze TIS-positive sequences that have their predictions changed to TIS-negative with the introduction of a single point mutation. Biologically, this means that, due to the introduced mutation, translation either fails or does not begin and the protein is not properly synthesized.


With the above process, we create $717,000$ mutated sequences originating from $1,195$ initially correctly classified TIS-positive sequences. Of those mutated sequences, we observe that $18,682$ have their predictions changed from TIS-positive to TIS-negative, with mutations causing this change.



In Figure~\ref{fig:mut_nuc_origin}, we provide the histogram of mutations where the mutation of base pairs to other bases in the graph cause a TIS-positive sequence to be classified as a TIS-negative. Based on results presented in Figure~\ref{fig:mut_nuc_origin}, we can align certain patterns with well-known biological features. 

\textbf{Importance of Kozak sequence}\,\textendash\,For humans as well as other higher eukaryotes, the sequence nearby the start codon is determined to play a significant role in translation initiation, with this (Kozak) sequence defined as 5'- \texttt{GCCRCC\blu{ATG}G} -3' with \texttt{R} representing a purine (i.e., \texttt{A} or \texttt{G}) and \blu{\texttt{ATG}} representing the start codon~\cite{kozak_structure}. It was discovered that the most influential base pairs in this sequence are the ones in position -3 (\texttt{R}) and +4 (\texttt{G}) relative to the start codon. Our experiments indicate that mutations made in these positions are also the ones that impact the prediction the most, with mutations on position -3 and +4 (often to \texttt{T}, which is the least occurring base pair in Kozak sequence) having the largest influence over prediction.


\textbf{Premature stop codons}\,\textendash\,The final stage of translation is termination, which signifies the end of the elongation of the peptide-chain, with this process requiring an in-frame stop codon (\texttt{TAA}, \texttt{TGA}, \texttt{TAG}). Premature termination of translation due to the introduction of a non-sense mutation often leads to specialized processes which prevent the creation of faulty proteins~\cite{johnson2011stops}. Our experiments indicate that most of the mutations that lead to TIS-negative prediction are the ones that create a stop codon and thus lead to a non-sense mutation. We provide the proportion of non-sense mutations to others in Figure~\ref{fig:mut_nuc_origin} where it is clear that these mutations mostly cause strong TIS-negative signals downstream of the start codon, as opposed to upstream.


\vspace{-0.5em}
\subsection{Attributions}
\label{sec:attributions}

The purpose of attribution methods is to identify important regions of the input data under consideration, where the information contained in these regions leads the examined DNN to make the type of prediction it makes~\cite{guided_backprop}. Outputs of attribution methods are often called attribution maps, and contain a measure of relative importance for each input location (e.g., pixel positions for images or nucleotide positions in genomic data).


By employing the seven attribution methods given in Section~\ref{sec:introduction} and by utilizing the $18,682$ mutated sequences that result in a TIS-negative prediction, we create $130,774$ attribution maps. Based on these attribution maps we identify the most important bp position as the location of largest attribution.



\begin{table}[t!]
\centering
\scriptsize
\caption{Percentage of attribution maps created with methods in the first row that identify mutations which occur in various regions of the sequence using the methodology described in Section~\ref{sec:attributions}.}
\label{tbl:attribution_match}
\begin{tabular}{clcccccccc}
\cmidrule[1pt]{1-9}
\multirow{2}{*}{\shortstack{Match}} & \multirow{2}{*}{\shortstack{Region}} & \multirow{2}{*}{\shortstack{DeepLift}} & Integrated & \multirow{2}{*}{\shortstack{LRP}} & Gradient & Guided & \multirow{2}{*}{\shortstack{Deconvolution}}  & Kernel \\
~ & ~ & ~ & Gradients & ~ & Shap & Backprop & ~ & Shap \\ 
\cmidrule[0.5pt]{1-9}
\multirow{10}{*}{\rotatebox[origin=c]{90}{Exact}} & Anywhere & 41.9\% & 44.3\% & \textbf{45.1}\% & 36.4\% & 42.0\% & 27.8\% & 7.0\% \\ 
~ & $\cdot$ Kozak & 32.8\% & 46.0\% & \textbf{46.7}\% & 43.1\% & 37.4\% & 27.2\% & 13.3\% \\ 
~ & $\cdot$ Upstream & 20.1\% & 28.0\% & \textbf{29.6}\% & 25.0\% & 22.7\% & 18.7\% & 7.3\% \\ 
~ & $\cdot$ Downstream & 49.1\% & 49.6\% & \textbf{50.2}\% & 40.2\% & 48.3\% & 30.8\% & 6.9\% \\ 
~ & $\cdot$ $\cdot$ Stop codons & 96.7\% & 96.5\% & \textbf{97.9}\% & 78.1\% & 96.6\% & 60.2\% & 11.4\% \\ 
~ & $\cdot$ $\cdot$ $\cdot$  \texttt{TAA} & 96.5\% & 96.4\% & \textbf{97.9}\% & 77.1\% & 96.5\% & 63.6\% & 10.3\% \\ 
~ & $\cdot$ $\cdot$ $\cdot$  \texttt{TAG} & 97.4\% & 97.2\% & \textbf{98.3}\% & 79.5\% & 96.9\% & 59.4\% & 12.1\% \\ 
~ & $\cdot$ $\cdot$ $\cdot$  \texttt{TGA} & 95.3\% & 95.2\% & \textbf{97.0}\% & 76.2\% & 96.0\% & 58.4\% & 10.8\% \\ 
~ & $\cdot$ $\cdot$  Other & 2.3\% & \textbf{3.6}\% & 3.3\% & 3.0\% & 1.0\% & 1.9\% & 2.5\% \\ 
\cmidrule[1pt]{1-9}
\end{tabular}
\vspace{-1.5em}
\end{table}

\textbf{Mutation identification by attribution methods}\,\textendash\,Given the nature of the experimental routine discussed in Section~\ref{sec:mutations}, we were able to identify the position that contributes the most to a TIS-negative prediction made by the model. Now, we investigate whether or not attribution methods can identify the positions of those mutations. Note that, although mutations occur at the DNA level, their effects may become apparent at the amino-acid level, which is the reason for stop codons having particular effects as discussed above. As a result of this, we evaluate the match for attribution on the level of codons, meaning that, if the attribution map highlights any of the three base pair of the mutated codon, we consider it a correct match. Furthermore, for mutations occurring in the Kozak sequence, we also consider an attribution to be a correct one if it highlights any region of the Kozak sequence.

Based on the evaluation routine described above, we provide Table~\ref{tbl:attribution_match} which contains the  percentage of attribution maps that are able to identify mutated codons for different regions of the sequence. Surprisingly, we observe that attribution methods have large discrepancies among each other when identifying appropriate regions, with \texttt{LRP} being the most accurate attribution method in identifying mutated codons. Furthermore, we also observe that stop codons that cause a premature termination of translation can be identified with a surprising accuracy ($+97\%$) even though these mutations are spread all over the downstream area (see Figure~\ref{fig:mut_nuc_origin}). Yet, for other mutations that occur in the downstream region, the accuracy of attribution shows a dramatic drop.

\vspace{-1em}
\section{Conclusions}
\vspace{-0.75em}

Utilizing the point mutations which see common use both in vivo and in vitro experiments, in this work, we performed a quantitative evaluation of attribution methods for DNN interpretability. We identified \texttt{LRP} to be the most accurate attribution method, with this method identifying biologically relevant features with surprising accuracy, yet still failing for half of the mutation scenarios, which raises the following question: why can attribution methods identify non-sense mutations in downstream regions extremely accurately, yet fail to identify other types of mutations? Supporting the observations of~\cite{koo2019robust} we also believe the answer lies in the evaluation of multiple models with varying degrees of performance in order to measure the correlation between the performance of the model and its ability to identify mutations. Nevertheless, we demonstrated that attribution methods, while being imperfect, can highlight regions relevant to the prediction for translation initiation and urge future research efforts to employ biologically significant experimentations for other datasets as well as for other models in order to discover faithfulness of proposed methods.

\bibliographystyle{unsrtnat}
\bibliography{refs}

\clearpage

\appendix

\section*{\huge{Appendix}}

\textbf{Data}\,\textendash\,An overview of the datasets used in this study is given in Table~\ref{tbl:dataset info} which details the positive to negative ratio of the labels as well as total number of sequences that fall into each category.

\textbf{Pre-processing}\,\textendash\,We use the standard one-hot-encoding routine for genomic data as described in the work of~\cite{zuallaert2017interpretable,zuallaert2018tisrover} which converts the genomic representation from strings to vectors as follows:
\begin{eqnarray}
\label{eq:data_desc}
\bm{x} = [ \, \underbrace{\mathtt{N}_{-60}\, \mathtt{N}_{-59}\, \ldots\, \mathtt{N}_{-2}\, \mathtt{N}_{-1}}_\text{Upstream}\,\,\, \mathtt{ATG}\,\,\, \underbrace{\mathtt{N}_{+4}\, \mathtt{N}_{+5}\, \ldots\, \mathtt{N}_{+142}\, \mathtt{N}_{+143}}_\text{Downstream} \, ]\, \,.
\end{eqnarray}
with $\texttt{N}_k \in \{[1, 0, 0, 0], [0, 1, 0, 0], [0, 0, 1, 0], [0, 0, 0, 1]\}$ for \texttt{A}, \texttt{C}, \texttt{T}, and \texttt{G}, respectively. We also employ with masking as an augmentation technique in order to improve the accuracy of the model where we replace one-hot representations of bp with empty representations. This operation corresponds to replacing the above vector depictions with a zero vector (i.e., $[0, 0, 0, 0])$.

\textbf{Model performance}\,\textendash\,~\cite{saeys2007translation} argues that commonly used metrics such as accuracy, true positive rate, and false positive rate can be misleading for datasets with an extreme skew in labels such as \texttt{Chromosome-21} and proposed the usage of false-positive rate at fixed sensitivity of $0.8$ (FPR.80) for benchmarking in TIS-detection. For easy comparability to the previous research efforts, we also use FPR.80 as an error measurement and provide the results of the best-performing models in Table~\ref{tbl:model info}. 

\begin{table}[h!]
\centering
\scriptsize
\begin{tabular}{lccccc}
\cmidrule[1pt]{1-6}
Dataset & Total & TIS-positive sequences & TIS-negative sequences & Positive/Negatie ratio & Source \\
\cmidrule[0.5pt]{1-6}
\texttt{CCDS} & 364,495 & 13,917 & 350,578 & 0.0396 & \cite{saeys2007translation} \\
\texttt{Chromosome-21} & 1,267,701 & 258 & 1,267,443 & 0.0002 & \cite{saeys2007translation} \\
\texttt{Human-TIS} & 2,718 & 1,359 & 1,359 & 1 & \cite{chen2014itis} \\
\cmidrule[1pt]{1-6}
\end{tabular}
\caption{Details of TIS-detection datasets used in this study.}
\label{tbl:dataset info}
\end{table}

\begin{table}[h!]
\centering
\begin{tabular}{lc}
\cmidrule[1pt]{1-2}
Model & FPR.80 \\
\cmidrule[0.5pt]{1-2}
\citet{saeys2007translation}  & 0.125 \\
TIS-Rover (reported in \citep{zuallaert2017interpretable}) & 0.031 \\
TIS-Rover (reproduced) & 0.032 \\
TIS-Rover (trained with SGD) & 0.030 \\
TIS-Rover (SGD + bp masking) & 0.029 \\
\cmidrule[1pt]{1-2}
\end{tabular}
\caption{FPR.80 performance of models on \texttt{Chromosome-21} dataset.}
\label{tbl:model info}
\end{table}

\textbf{Distance between attributions and mutations}\,\textendash\,In the main text, in order to measure the correctness of attribution maps, we measured the whether or not the largest attribution location lies within one of the three bp of the codon that is mutated. In Table~\ref{tbl:attribution_match2}, we provide experimental results for showing mean and median codon distance between the attributions and mutations for the same regions. With the results displayed in Table~\ref{tbl:attribution_match2}, we show that, for mutations that do not occur in the Kozak sequence or the ones that do not create a stop codon in downstream region, attribution methods highlight regions that are, on average, far away from the mutations.

\begin{table*}[t!]
\centering
\caption{Mean (median) distance between the codon highlighted by attribution maps created with methods in the first row and codons that are mutated. The results are provided separately for a number of regions in the second column.}
\scriptsize
\begin{tabular}{clcccccccc}
\cmidrule[1pt]{1-9}
\multirow{2}{*}{\shortstack{Match}} & \multirow{2}{*}{\shortstack{Region}} & \multirow{2}{*}{\shortstack{DeepLift}} & Integrated & \multirow{2}{*}{\shortstack{LRP}} & Gradient & Guided & \multirow{2}{*}{\shortstack{Deconvolution}}  & Kernel \\
~ & ~ & ~ & Gradients & ~ & Shap & Backprop & ~ & Shap \\ 
\cmidrule[0.5pt]{1-9}
\multirow{10}{*}{\rotatebox[origin=c]{90}{Distance to mutation}} & Anywhere & 9.3 (3) & 8.7 (2) & 8.9 (2) & 9.9 (5) & 10.8 (3) & 17.3 (13) & 17.2 (15) \\ 
~ & $\cdot$ Kozak & 10.3 (7) & 6.9 (0) & 8.4 (4) & 6.8 (0) & 10.6 (7) & 13.0 (16) & 14.3 (11) \\ 
~ & $\cdot$ Upstream & 14.2 (11) & 10.5 (8) & 11.1 (7) & 10.4 (8) & 13.1 (9) & 10.2 (9) & 16.6 (12) \\ 
~ & $\cdot$ Downstream & 7.6 (1) & 8.1 (1) & 8.1 (0) & 9.8 (3) & 10.0 (1) & 19.6 (19) & 17.4 (16) \\ 
~ & $\cdot$ $\cdot$ Stop codons & 1.0 (0) & 1.2 (0) & 0.8 (0) & 3.9 (0) & 1.5 (0) & 12.1 (0) & 16.1 (14) \\ 
~ & $\cdot$ $\cdot$ $\cdot$ \texttt{TAA} & 1.2 (0) & 1.3 (0) & 0.9 (0) & 4.0 (0) & 1.6 (0) & 11.9 (0) & 16.5 (14) \\ 
~ & $\cdot$ $\cdot$ $\cdot$ \texttt{TAG} & 0.7 (0) & 0.9 (0) & 0.7 (0) & 3.6 (0) & 1.4 (0) & 11.4 (0) & 15.9 (14) \\ 
~ & $\cdot$ $\cdot$ $\cdot$ \texttt{TGA} & 1.4 (0) & 1.8 (0) & 1.2 (0) & 4.6 (0) & 1.7 (0) & 13.5 (0) & 16.1 (14) \\ 
~ & $\cdot$ $\cdot$ Other & 14.7 (13) & 15.4 (13) & 15.9 (14) & 16.0 (14) & 19.0 (17) & 27.6 (28) & 18.7 (17) \\ 
\cmidrule[1pt]{1-9}
\end{tabular}
\label{tbl:attribution_match2}
\end{table*}

\end{document}